\documentclass[11pt,a4paper]{article}
\usepackage{amsmath}
\usepackage{amsfonts}
\usepackage[left=2.50cm, right=2.50cm, top=3.00cm, bottom=3.00cm]{geometry}
\usepackage{amssymb}
\usepackage{graphicx}
\usepackage[space]{grffile}
\usepackage{xcolor}
\usepackage{comment}
\usepackage{newtxtext}
\usepackage{newtxmath}
\usepackage{natbib}
\usepackage{hyperref}
\hypersetup{
    colorlinks = true,
    urlcolor   = blue,
    citecolor  = black,
}

\author{David Andrade \& Raphael Stuhlmeier}
\date{Centre for Mathematical Sciences, University of Plymouth, Drake Circus, PL4 8AA, Plymouth, UK}
\title{The nonlinear Benjamin-Feir instability -- Hamiltonian dynamics, primitive breathers, and steady solutions}

\newcommand{\bk}{\mathbf{k}}

\begin{document}

\maketitle

\begin{abstract}
We develop a general framework to describe the cubically nonlinear interaction of a unidirectional degenerate quartet of deep-water gravity waves. Starting from the discretised Zakharov equation, and thus without restriction on spectral bandwidth, we derive a planar Hamiltonian system in terms of the dynamic phase and a modal amplitude. This is characterised by two free parameters: the wave action and the mode separation between the carrier and the side-bands. The mode separation serves as a bifurcation parameter, which allows us to fully classify the dynamics. Centres of our system correspond to non-trivial, steady-state nearly-resonant degenerate quartets. The existence of saddle-points is connected to the instability of uniform and bichromatic wave trains, generalising the classical picture of the Benjamin-Feir instability. Moreover, heteroclinic orbits are found to correspond to primitive, three-mode breather solutions, including an analogue of the famed Akhmediev breather solution of the nonlinear Schr\"odinger equation.   
\end{abstract}


\section{Introduction}

The linear theory of water waves, developed and refined over the course of the 19$^{\text{th}}$ and early 20$^{\text{th}}$ centuries has endured by virtue of its simplicity, elegance, and utility. When the governing equations are linearised, waves may be superposed to create the myriad complex patterns observed on the surface of the sea. The absence of interaction among such waves leads, however, to an inability to explain certain properties of ocean wave fields, particularly if these waves are steep or the time-scales considered are long.

This interaction may occur between a wave and itself, as discovered by G.\ G.\ Stokes in \citeyear{Stokes1847}, or as a mutual interaction between different waves. A natural way to think about such weakly-nonlinear interactions is as occurring between Fourier modes, each of which represents a periodic wave train (or plane wave) with a distinct frequency, amplitude, and phase. When considering gravity waves in deep water, the fundamental interaction is between four such Fourier modes, which may or may not be distinct. This fact is not obvious a priori, as it depends on the possibility of resonance between modes, and therefore on the dispersion relation of the problem. The deep-water dispersion relation $\omega^2 = g |\mathbf{k}|$ between radian frequency $\omega$ and wavenumber $\mathbf{k}$ provides a geometric constraint which means resonance is possible for cubic (or higher) nonlinearity. This pioneering discovery by \cite{Phillips1960} soon became the object of intense study, with work to elucidate its consequences for everything from wave spectra \citep{Hasselmann1962a} to the evolution of uniform wave trains (so-called Stokes waves) \citep{Benjamin1967a,Zakharov1968}.

One of the most remarkable early results was the finding that initially uniform wave-trains are unstable, and tend to disintegrate into wave groups -- known as modulational or Benjamin-Feir instability after \cite{Benjamin1967a}. Indeed, uniform wave-trains are often difficult to produce experimentally, a fact  attributed to this instability. The perturbations required to start the process of disintegration are small-amplitude side-bands, or Fourier modes to either side of the plane wave. Interaction between the modes transfers energy to the side-bands, which grow at the expense of the plane wave (or carrier) and distort the pattern on the free-surface.  Shortly after the discovery of the instability \cite{Zakharov1968} was able to reduce the problem to Hamiltonian form, and to derive a simplified equation governing the evolution of such wave groups for unidirectional waves in deep water, provided the wave modes are not widely separated in Fourier space. This nonlinear Schr\"{o}dinger equation (NLS) was then used to recover Benjamin and Feir's criterion, specifying which ranges of side-band wavenumbers and carrier steepness give rise to instability. 

The nonlinear Schr\"odinger equation has also received considerable recent attention in connection with the deterministic and stochastic modelling of extreme waves, large wave crests which appear without warning on the surface of the sea (a recent review of progress is given by \cite{Onorato2016a}). The temporally or spatially localised exact solutions of the NLS, called ``breathers,'' are of particular importance to our understanding of extreme waves. Their recent experimental observation \citep{Chabchoub2014} make them an attractive model to generate extreme waves in the wave flume, and shed light on the behaviour of such waves in the ocean. Indeed, each breather is a particular manifestation of the underlying Benjamin-Feir instability for special initial conditions.

The success of the NLS as a simplified model equation naturally engendered interest in its agreement with more general water wave equations. The NLS breather solutions, for example, were found to be robust when compared to numerical solutions of the Euler equations by \cite{Slunyaev2013}, who also gave an overview of previous work in this vein. The Benjamin-Feir stability threshold derived from the NLS has also been re-examined by \cite{Crawford1981}, who sought to remove the restriction to small mode separation. However, the route to obtaining an instability criterion in their work remained one of linearisation and the derivation of an eigenvalue criterion. The smallness assumptions intrinsic to linear stability analysis mean that it is limited to describing only the initial stages of evolution around particular exact solutions. In this setting, the eigenvalues of the linearised system govern the growth rate of initially small disturbances, whose subsequent evolution can be captured by numerical simulation. However, a numerical treatment has the potential to obscure fundamental, underlying mechanisms, particularly in cases with many (ostensibly independent) parameters. For example, computing the evolution of a carrier and two side-bands requires the specification of eight parameters: two wavenumbers $\bk_a, \bk_b,$ (the third is given from the resonance condition $2\bk_a = \bk_b + \bk_c$), three wave slopes $ \epsilon_a, \epsilon_b, \epsilon_c,$ and three phases $\theta_a, \theta_b,$ and $\theta_c.$

We aim to present a unified theory of the cubically nonlinear interaction of three waves in deep water, which generalises the Benjamin-Feir instability. Our starting point, like that of \cite{Crawford1981}, is the reduced Zakharov equation (ZE), which is free from any restrictions on mode separation. Indeed, the nonlinear Schr\"odinger equation itself, as well as later generalisations due to \cite{Societyd} or \cite{Dysthe1979} can all be derived as narrow-band limits of the ZE (see \cite{Stiassnie1984c,GRAMSTAD2011}). After suitable transformations, the interaction of three wave modes satisfying $2\bk_a = \bk_b + \bk_c$ can be recast as a planar Hamiltonian dynamical system. Phase plane analysis of this system allows for a complete description of the dynamics for all times and arbitrary initial conditions: these initial conditions are reduced to specifying a single ``dynamic phase'' and a parameter specifying the distribution of the wave action among the three modes. The mode-separation distance plays the role of a bifurcation parameter in the problem.

With this simplification we are able to fully classify the dynamics of such ``degenerate quartets'' of unidirectional waves. Fixed points in the phase-plane correspond to steady-state near-resonant cases, and heteroclinic orbits are seen to correspond to a variety of primitive breather-type solutions. We obtain general instability results for uniform and bichromatic wave trains as criteria for the existence of fixed-points, and show that the well-known results of linear stability analysis can be recovered. In what follows we first present the reformulation of the discrete Zakharov equation for a degenerate quartet of waves in section \ref{sec:ZE for degenerate quartet}. In section \ref{sec:Hamiltonian dynamics} we classify the fixed points and phase-portraits, using mode separation as a bifurcation parameter. The nonlinear stability of wave trains is considered in section \ref{sec:Nonlinear stability}, and special, heteroclinic solutions are discussed in section \ref{sec:Heteroclinic solutions}. Finally, a discussion of our results is presented in section \ref{sec:Discussion}. Appendix \ref{app:kernel} contains some simplified expressions for integral kernels used in computations.

\section{The Zakharov equation for a degenerate quartet of waves}
\label{sec:ZE for degenerate quartet}

Investigating third-order wave-wave interaction on deep water without a bandwidth restriction means that our starting point will be the reduced Zakharov equation
\begin{equation} \label{eq: Discrete Zakharov Equation}
i \frac{dB_n}{dt} = \sum_{p,q,r =1}^N T_{npqr} \delta_{np}^{qr} e^{i \Delta_{npqr} t} B_p^* B_q B_r, \, n = 1, \, 2, \ldots , N
\end{equation}
derived by Zakharov \citeyearpar{Zakharov1968}, and in Hamiltonian form by Krasitskii \citeyearpar{Krasitskii1994}, and here written in a convenient discrete formulation. While the equation itself has the generic form of a cubic interaction equation (see \cite{Zakharov1992a}), the physics of the water-wave problem are contained in the kernels  $T_{npqr}=T(\bk_n,\bk_p,\bk_q,\bk_r).$ We use $\delta_{np}^{qr}$ to denote a Kronecker delta function 
\begin{equation*}
\delta_{np}^{qr} = \begin{cases} 1 \text{ for } \bk_n+\bk_p=\bk_q+\bk_r \\
0 \text{ else.} \end{cases}
\end{equation*}
and $\Delta_{npqr} = \omega_n + \omega_p - \omega_q - \omega_r$ to denote the frequency detuning (a measure of departure from exact resonance). Throughout 
\begin{equation} \label{eq: Linear Dispersion Relation}
\omega_n^2 = g | \bk_n|
\end{equation}
is the linear dispersion relation for gravity waves in deep water.

The simplest non-trivial interaction is between three waves in deep water $\bk_a, \bk_b$ and $\bk_c,$ where one wave is counted twice to satisfy the resonance condition $2\bk_a = \bk_b + \bk_c.$ Indeed this case -- termed the ``degenerate quartet'' -- corresponds to the Benjamin-Feir instability (see \cite{Yuen1982} or \cite[Ch.\ 14.9]{Mei2005}), where mode $\bk_a$ is interpreted as a ``carrier wave'' and modes $\bk_b$ and $\bk_c$ as ``side-bands.''  

Rather than handle three complex equations, it is a convenient first step towards simplification to write the discrete equation in terms of amplitude and phase variables (see, e.g.\ \cite{Craik1986} and references therein). The equations then become 
\begin{subequations} \label{eq:3 Wave-ZE}
\begin{align} \label{eq:DQ-Ba}
\frac{d |B_a|}{dt} &= 2 T_{aabc} |B_a||B_b||B_c| \sin(\theta_{aabc}), \\ \label{eq:DQ-Bb}
\frac{d |B_b|}{dt} &= - T_{aabc} |B_a|^2|B_c| \sin(\theta_{aabc}), \\\label{eq:DQ-Bc}
\frac{d |B_c|}{dt} &= - T_{aabc} |B_a|^2|B_b| \sin(\theta_{aabc}), \\ 
\frac{d\theta_a}{dt} &= \Gamma_a + 2\frac{T_{aabc}}{|B_a|^2}|B_a|^2|B_b||B_c|\cos(\theta_{aabc}),\label{eq:DQ_Theta_a}\\
\frac{d\theta_b}{dt} &= \Gamma_b + \frac{T_{aabc}}{|B_b|^2}|B_a|^2|B_b||B_c|\cos(\theta_{aabc}),\label{eq:DQ_Theta_b}\\
\frac{d\theta_c}{dt} &= \Gamma_c + \frac{T_{aabc}}{|B_c|^2}|B_a|^2|B_b||B_c|\cos(\theta_{aabc}),\label{eq:DQ_Theta_c}
\end{align}
\end{subequations}
where $\theta_{aabc} = \Delta_{aabc}t + 2\theta_a - \theta_b - \theta_c$ is called the dynamic phase.
Note the additional factor of two appearing in the equations for $|B_a|$ and $\theta_a$ because the resonance condition $2\bk_a = \bk_b + \bk_c$ can be satisfied for either the tuple $(\bk_a,\bk_a,\bk_b,\bk_c)$ or $(\bk_a,\bk_a,\bk_c,\bk_b)$.

In equations \eqref{eq:DQ_Theta_a} - \eqref{eq:DQ_Theta_c} 
\begin{equation}\label{eq: Gamma}
\Gamma_i = |B_i|^2T_i + 2 \sum_{j\neq i} |B_j|^2 T_{ij},
\end{equation}
where we use an abbreviated notation for the symmetric kernels: $T_i = T_{iiii}, \, T_{ij}=T_{ijij}=T_{jiji}$. 

The next significant simplification relies on the observation that, although there are ostensibly three distinct phases in the equations governing the degenerate quartet \eqref{eq:DQ-Ba}--\eqref{eq:DQ_Theta_c}, they occur only in the single combination $2\theta_a - \theta_b - \theta_c,$ which makes it possible to drop subscripts and write $\theta$ and $\Delta$ in place of $\theta_{aabc}$ and $\Delta_{aabc}$ without risk of confusion. We can further exploit this fact to combine \eqref{eq:DQ_Theta_a} - \eqref{eq:DQ_Theta_c} into a single equation for the dynamic phase
\begin{align}\label{eq:DQ-Theta}
\frac{d \theta}{dt} &= \Delta + 2 \Gamma_a - \Gamma_b - \Gamma_c + T_{aabc} \left( 4 |B_b||B_c| - \frac{|B_c||B_a|^2}{|B_b|} - \frac{|B_b||B_a|^2}{|B_c|}\right) \cos(\theta).
\end{align}

Equations \eqref{eq:DQ-Ba} - \eqref{eq:DQ-Bc} and \eqref{eq:DQ-Theta} now form an autonomous system of ordinary differential equations. It is known that this system of equations admits periodic solutions which are given in terms of Jacobi elliptic functions, see \cite{Shemer1985}. We shall see that generic solutions are indeed periodic, though we shall focus our attention principally on special, non-periodic solutions. 

The system of equations \eqref{eq:DQ-Ba}-\eqref{eq:DQ-Bc} and \eqref{eq:DQ-Theta} admits the following conserved quantities:
\begin{align} \label{eq:Conservation-Wave Action}
|B_a|^2 + |B_b|^2 + |B_c|^2 = A,
\end{align}
the total wave action, and 
\begin{align} \label{eq:Conservation-alpha}
\frac{|B_b|^2}{A} - \frac{|B_c|^2}{A} = \alpha,
\end{align}
the wave-action difference in the two side-bands. These conserved quantities allow us to further reduce the number of parameters, by introducing a normalized wave-action variable $\eta = |B_a|^2/A$, which by \eqref{eq:Conservation-Wave Action} must take values between $0$ and $1$. The amplitudes may then be rewritten in terms of $\eta$ and $\alpha$ as $|B_a|^2 = A\eta$, $|B_b|^2 = A(1 - \eta + \alpha)/2$ and $|B_c|^2 = A(1 - \eta - \alpha)/2$.

In terms of the two variables $\eta$ and $\theta$ we find that the system can be described by the Hamiltonian 
\begin{align}\label{eq:H}
H(\theta,\eta) = -AT_{aabc}2\eta\sqrt{(1 - \eta)^2 - \alpha^2}\cos(\theta) - (\Delta + A\Omega_0)\eta - \frac{A\Omega_1}{2}\eta^2,
\end{align}
with
\begin{align}
\frac{\partial H}{\partial \theta} &= \frac{d\eta}{dt} = 2AT_{aabc}\eta\sqrt{(1 - \eta)^2 - \alpha^2}\sin(\theta),  \label{eq:eta:1} \\
-\frac{\partial H}{\partial\eta} &= \frac{d\theta}{dt} = \Delta_{aabc} + A\Omega_0 + A\Omega_1\eta + 2AT_{aabc}\frac{1 + 2\eta^2 - 3\eta - \alpha^2}{\sqrt{(1 - \eta)^2 - \alpha^2}}\cos(\theta).\label{eq:theta:1}
\end{align}
The two new coefficients are
\begin{align}
    \label{eq:Omega0}
    \Omega_0 &= 2 \left[ T_{ab}(1+\alpha) + T_{ac}(1-\alpha) \right] - \frac{1}{2}\left[ T_b (1+\alpha) + T_c (1-\alpha) \right] - 2T_{bc}\\
    \Omega_1 &= 2 \left( T_a + T_{bc} \right) - 4 \left( T_{ab} + T_{ac} \right) + \frac{1}{2} \left( T_b + T_c \right). \label{eq:Omega1}
\end{align}
Such a Hamiltonian approach to the three-wave discretisation of the nonlinear Schr\"odinger equation was first explored by \cite{Cappellini1991} in the context of optics, and our Hamiltonian \eqref{eq:H} can be shown to reduce in the narrow-band limit to one analogous to that presented therein.

A final simplification of the system can be achieved by imposing an equidistribution of wave action among the side-bands, i.e.\ $\alpha = 0.$ Under this assumption equations \eqref{eq:eta:1} and \eqref{eq:theta:1}, as well as the Hamiltonian \eqref{eq:H}, become 
\begin{align}
\frac{d\eta}{dt} &= 2AT_{aabc}\eta(1 - \eta)\sin(\theta),\label{eq:eta:alpha=0}\\\label{eq:theta:alpha=0}
\frac{d\theta}{dt} &= \Delta + A\Omega_0 + A\Omega_1\eta + 2AT_{aabc}(1 - 2\eta)\cos(\theta),\\
H(\theta,\eta) &= -2AT_{aabc}\eta(1 - \eta)\cos(\theta) - (\Delta + A\Omega_0)\eta -\frac{A\Omega_1}{2}\eta^2.\label{eq:H:alpha=0}
\end{align} 

In subsequent computations we shall normalise all wavenumbers by the carrier wavenumber $k_a,$ and all frequencies by the carrier frequency $\omega_a$, which is equivalent to setting the gravitational acceleration $g = 1$. For the unidirectional cases considered we shall also explicitly specify our degenerate quartets in terms of mode separation $p$ as follows:  $\bk_a = [1,0]$, $\bk_b = [1 - p,0]$ and $\bk_c = [1 + p,0]$, where $0<p<1.$ We henceforth drop the bold-script and write $k_i$ for the wavenumbers, to emphasise that these are scalars. Indeed, for unidirectional waves in deep water it is possible to significantly simplify the kernels, as shown by \cite{Dyachenko2017,Kachulin2019} and detailed in Appendix \ref{app:kernel}.

The wave action $A$ of the three-mode system remains a free parameter, and the conservation law \eqref{eq:Conservation-Wave Action} makes it clear that $A$ may be ``distributed" among the three modes in various configurations. The relationship between the complex amplitudes $B_i$ of the Zakharov formulation and the amplitude of the free surface displacement (e.g.\ (14.5.5) of \cite[Sec.\ 14.5]{Mei2005}) makes it possible to interpret $A$ in terms of the carrier steepness, which is sometimes convenient. If only the carrier wave is present
\begin{equation} \label{eq:A-carrier-rel}
A = |B_a|^2 = \frac{2 g a^2 \pi^2}{\omega_a} = \frac{2 \pi^2 g^{1/2} \epsilon^2}{k_a^{5/2}}, 
\end{equation}
where $a$ is the surface-wave amplitude and $\epsilon=a k_a$ is the wave slope.

Aside from the wave action $A,$ the system \eqref{eq:eta:alpha=0} and \eqref{eq:theta:alpha=0} contains a further free parameter: the separation between the Fourier modes $p.$ Together these can be used to determine the values of all coefficients in the equation.

\section{Hamiltonian dynamics in the phase-plane}
\label{sec:Hamiltonian dynamics}

The Hamiltonian dynamical system \eqref{eq:eta:alpha=0} -- \eqref{eq:theta:alpha=0} describes fully the nonlinear dynamics of three interacting deep-water waves. We can gain both qualitative and quantitative understanding of this system via phase--plane analysis, where the phase space is the surface of the truncated cylinder $\{(\theta,\eta) \mid -\pi \leq \theta \leq \pi, \, 0\leq \eta \leq 1 \}.$

\subsection{Fixed points -- steady state degenerate quartets}
\label{ssec:Fixed points}

The first step in unraveling the phase dynamics is to consider fixed points of our system. The trajectories, corresponding to solutions of the degenerate quartet with particular initial conditions, occupy the level curves of the Hamiltonian \eqref{eq:H:alpha=0}. We find that fixed points of \eqref{eq:eta:alpha=0} -- \eqref{eq:theta:alpha=0} occur in four classes: $\eta = 0$ or 1 and $\theta = 0$ or $\pm\pi.$ The upper and lower boundaries $\eta=1$ and $\eta=0$ correspond to a single wave train and a bichromatic wave train, respectively (recall $\eta=|B_a|^2/A$). These two well-known solutions of the Zakharov equation exhibit no energy exchange (see \cite{Leblanc2009} or \citet[Sec.\ 14.5 \& 14.6]{Mei2005}), since $d\eta/dt = 0$. Indeed, the effect of nonlinear interaction in such cases is solely to induce a frequency correction (called Stokes' correction after Stokes \citeyearpar{Stokes1847}, and found for bichromatic waves in Longuet-Higgins \& Phillips \citeyearpar{Longuet-Higgins1962c}; for a discussion in the context of the Zakharov equation see Stuhlmeier \& Stiassnie \citeyearpar{Stuhlmeier2019}).

To avoid overly bulky expressions in what follows we introduce the following notation
\begin{align}
    \Delta' = \frac{\Delta}{AT_{aabc}}, \quad \Omega_0' = \frac{\Omega_0}{T_{aabc}}, \quad \text{ and }\quad \Omega_1' = \frac{\Omega_1}{T_{aabc}}.
\end{align}
In these variables we obtain fixed points at $\eta=1$  of the form $(\theta,\eta)=(\theta_{\pm 1},1)$ whenever $\theta_{ 1}$ satisfies
\begin{align}\label{fixpoint:eta1}
    \cos(\theta_1) = \frac{\Delta' + \Omega_0' + \Omega_1'}{2}.
\end{align}
Such fixed points exist when the right-hand side is between -1 and 1, and $\theta_{-1} = -\theta_1.$ These fixed points lie on the contour $H=-\Delta - A\Omega_0 - \frac{A\Omega_1}{2}.$ 

Similarly, we obtain fixed points at $\eta=0$ of the form $(\theta,\eta)=(\theta_{\pm 0},0)$ whenever $\theta_{ 0}$ satisfies
\begin{align} \label{fixpoint:eta0}
    \cos(\theta_0) = -\frac{\Delta' + \Omega_0'}{2},
\end{align}
again provided the right hand side is between -1 and 1 and $\theta_{-0} = -\theta_0$. These fixed points lie on the contour $H=0.$

Another class of fixed points is obtained for $\theta=0$ or $\pm \pi.$ In the former case $(\theta,\eta)=(0,\eta_0)$ is a fixed point whenever
\begin{align}\label{fixpoint:theta0}
\eta_0 = \frac{2 + \Delta' + \Omega_0'}{4 - \Omega_1'},
\end{align}
and $0\leq\eta_0 \leq 1$.
Note that $\eta_0 = 1$ if and only if $\theta_{\pm 1} = 0,$ in which case this fixed point coincides with \eqref{fixpoint:eta1} above. 
For $\theta =\pm\pi$ we find fixed points $(\pm\pi,\eta_\pi)$ provided 
\begin{align}\label{fixpoint:thetaPi}
\eta_\pi = \frac{2 - (\Delta' + \Omega_0')}{4 + \Omega_1'},
\end{align}
and $0\leq\eta_\pi \leq 1$. Similarly, $\eta_\pi = 0$ if and only if $\theta_{\pm 0} = \pm\pi,$ such that this fixed point coincides with \eqref{fixpoint:eta0}. 

\subsection{Phase portraits and bifurcation}
\label{ssec:Phase portraits and bifurcation}

The dynamical system \eqref{eq:eta:alpha=0} -- \eqref{eq:theta:alpha=0} contains two parameters -- wave action $A$ and mode-separation $p$ -- which together govern the existence of the fixed-points given in equations \eqref{fixpoint:eta1}--\eqref{fixpoint:thetaPi}, and the attendant dynamics. The natural choice is to fix $A$ -- akin to specifying the total energy of waves -- and to use mode separation as a bifurcation parameter. 

\begin{figure}
\centering
\includegraphics[width=\linewidth]{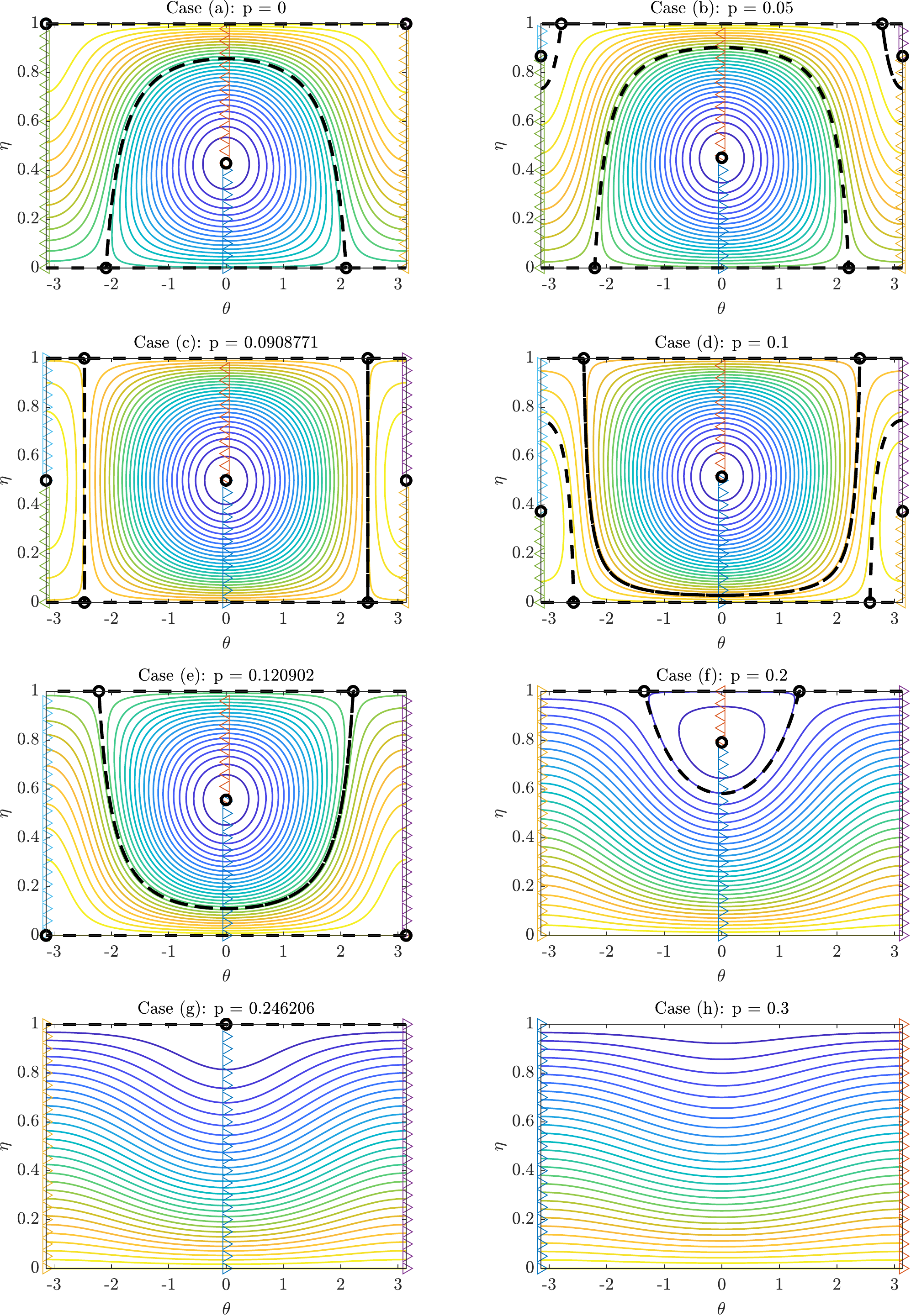}
\caption{Phase portraits for $A=2 \pi^2/100$ and various values of mode separation $p$ between 0 and 0.3, plotted on the cylinder $\{(\eta,\theta)\mid \eta \in [0,1], \theta \in [-\pi,\pi]\}.$ Coloured curves represent contours of the Hamiltonian \eqref{eq:H:alpha=0}. Dashed lines show separatrices, while dots denote fixed-points of the system.}
\label{fig:Phase portraits}
\end{figure}

In figure \ref{fig:Phase portraits} we depict a series of characteristic phase portraits, where mode separation $p$ increases from $p=0$ in Case (a) to $p=0.3$ in Case (h). In this figure we have selected $A$ equivalent to a carrier wave with steepness $\epsilon=0.1,$ so that $A=2 \pi^2/100$ (see \eqref{eq:A-carrier-rel} and recall $g=k_a=1$). The figure thus represents both physically realistic, as well as characteristic behaviour, in a sense to be specified in detail below.

Deferring the details for the moment we can give an overview of the dynamics of our degenerate quartet as obtained by phase-plane analysis, and shown in figure \ref{fig:Phase portraits}. In the limit of vanishing mode separation, Case (a), we find four fixed points: a centre at $\theta=0,$ a pair of saddle points on $\eta=0,$ and a semi-stable fixed point at $\theta=\pm \pi, \, \eta=1.$ The first bifurcation occurs at $p=0$, and as $p$ increases the semi-stable point at $(\theta,\eta)=(\pm \pi,1)$ splits into three fixed points: a centre at $\theta=\pm \pi$ and two saddles at $\eta=1$ (Case (b)). Further increase in the mode separation causes the centre at $\theta=\pm \pi$ to descend until the saddle connections between the two fixed points at $\eta=0$ and $\eta=1$ merge (Case (c)). 

Further increasing  $p$ leads the centre at $\theta=\pm \pi$ to descend towards $\eta=0$ (Case (d)) where it subsequently vanishes together with the saddle points along $\eta=0$ (Case (e)). As the modes are separated further the remaining three fixed points draw closer together in phase space (Case (f)) before coalescing (Case (g)) and disappearing entirely (Case (h)) -- this last bifurcation leads to complete stabilisation of the system. 

\section{Nonlinear stability}
\label{sec:Nonlinear stability}

While the phase-plane analysis presented above is remarkably simple it contains a wealth of information about the fully nonlinear dynamics of interacting degenerate quartets of deep water waves. We focus first on a discussion of nonlinear instability results. 

\subsection{Nonlinear instability of a uniform wave train}
\label{ssec:Nonlinear instability - uniform wave train}

Many classical studies of the stability of a uniform wave train begin by establishing that such a monochromatic wave is a solution of the relevant governing equation. This solution is then used as a starting point for linearisation, and an instability criterion derived from the eigenvalues of the linear system, as in \cite{Crawford1981}. In our framework, uniform wave trains are described by $\eta=1,$ for arbitrary values of $\theta.$ 

The initial small disturbance used classically to investigate the Benjamin-Feir instability consists in imposing small side-bands, i.e.\ a shift to a contour $\eta<1.$ In cases (a) through (f) of figure \ref{fig:Phase portraits} we readily observe that such a shift leads to growth of the side-bands, as $\eta$ decreases along the contours, followed generically by periodic energy exchange. Maximal energy exchange occurs when $\eta$ is changing fastest, and coincides with the smallest changes in the dynamic phase $\theta.$ Such phase coherence has also been observed in simulations, e.g.\ \cite{Houtani_2022,LiuWasedaZhang2021}.

In fact we can show that the classical linear stability criterion is equivalent to the existence of fixed points at $\eta = 1$ in the nonlinear system. The condition \eqref{fixpoint:eta1} for such fixed points to exist is 
\begin{align}
    -1 \leq \frac{\Delta' + \Omega_0' + \Omega_1'}{2}\leq 1.
\end{align}
Squaring these inequalities, substituting $\Omega_0$ and $\Omega_1$ from \eqref{eq:Omega0} and \eqref{eq:Omega1}, and using \eqref{eq:A-carrier-rel} yields, after simplification, the discriminant criterion 
\begin{align}\label{Discriminant}
    \left(\frac{\Delta}{2} + (T_a - T_{ab} - T_{ac})|B_a|^2\right)^2 - |B_a|^4T_{aabc}^2\leq 0,
\end{align}
see \citep[Eq.\ (14.9.16)]{Mei2005}. It is rather surprising that this eigenvalue condition, which arises from an approach based on small amplitude perturbations (and which is oblivious to the existence of fixed-points of the original nonlinear problem) can be recovered exactly with our approach.

\subsubsection{Restabilisation}
\label{sssec:Restabilisation}

The existence of fixed points at $\eta=1,$ and thus the instability of uniform wave trains to \textit{some} perturbation, is the generic situation for degenerate quartets. We can appreciate this by considering equation \eqref{fixpoint:eta1}, from which we establish that fixed points exist at $\eta=1$ for some mode separation, provided 
$ A \in \left[ 0 , {2 \pi (2-\sqrt{2}) \sqrt{g}}{k_a^{-5/2}} \right]. $
Employing \eqref{eq:A-carrier-rel} this can be expressed in terms of carrier wave slope $\epsilon,$ and implies the existence of fixed points (and thus instability) for 
\[ 0 < \epsilon < \sqrt{2-\sqrt{2}},\]
far beyond the wave breaking threshold.

\begin{figure}
    \centering
    \includegraphics[scale=0.5]{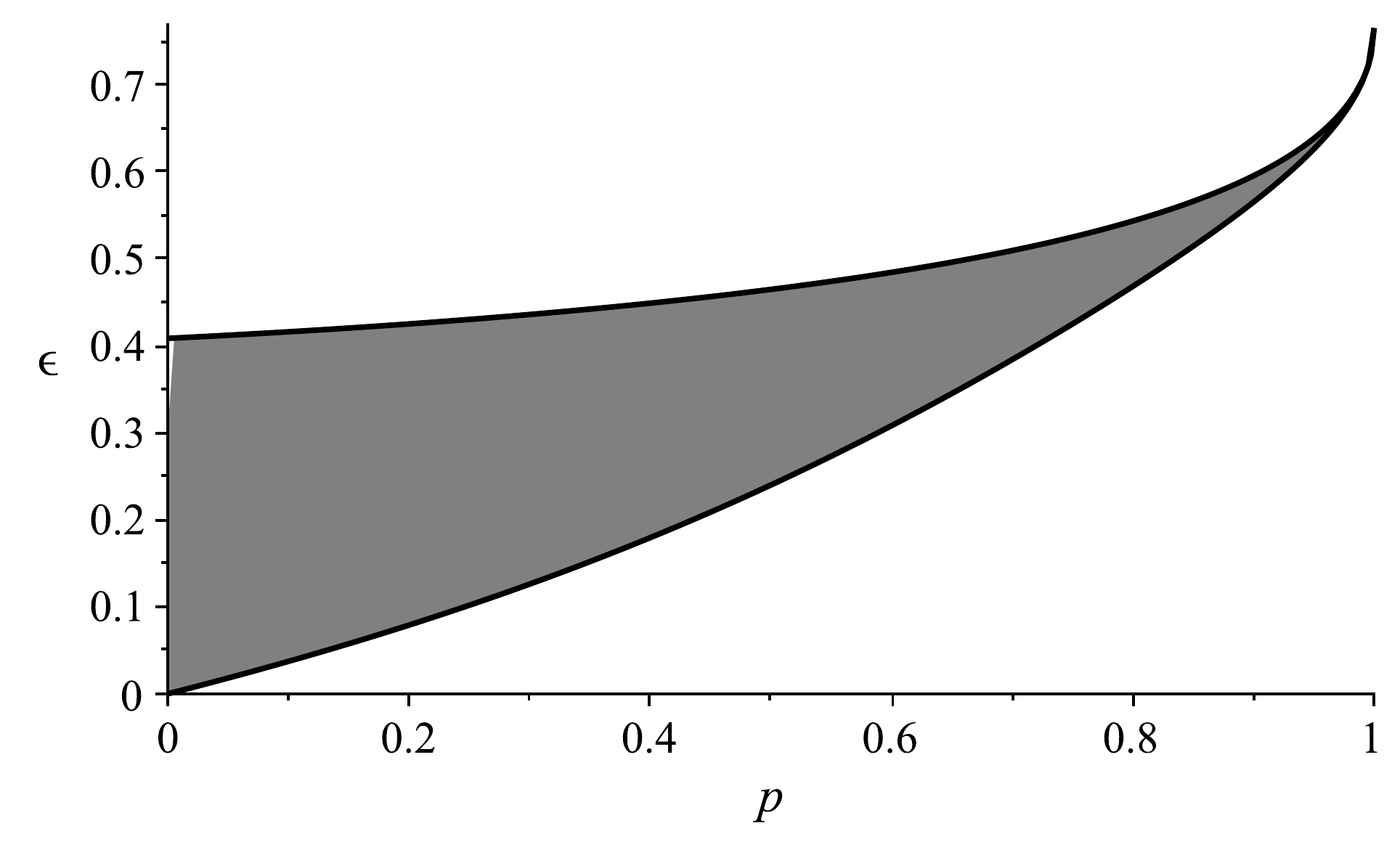}
    \caption{Existence of $\eta=1$ fixed points in $(\epsilon,p)$-parameter space according to \eqref{fixpoint:eta1}. The shaded region denotes the nonlinear instability domain for the degenerate quartet}
    \label{fig:eta_1 instability}
\end{figure}

In the limit of small mode separation, we see that the instability domain remains bounded (see figure \ref{fig:eta_1 instability}), with fixed points for $A \in [0,\frac{\sqrt{g k_a}\pi^2}{3 k_a}].$ This yields the threshold
\[ 0 < \epsilon < \frac{1}{\sqrt{6}}, \]
which has previously been obtained numerically (see e.g.\ \cite[Sec.\ VI.B.1]{Yuen1982}). In fact, this restabilisation for nearby side-bands (or long-wavelength disturbances) is a characteristic of the broad-banded Zakharov equation. Indeed, imposing a limit of narrow spectral bandwidth (in which the Zakharov equation reduces to the NLS) implies 
\[ \Omega'_0 \rightarrow 1, \, \Omega'_1 \rightarrow -3, \text{ and } \Delta' \rightarrow \frac{\omega_a p^2 \pi^2}{A k_a^5}.\]
This can be used to reformulate the fixed-point critrion and recover the well-known instability criterion for NLS \citep[Eq.\ (14.9.21)]{Mei2005}
\[ \frac{p}{k_a} \in \left[ 0,2\sqrt{2} \epsilon\right]. \]

As mode-separation increases, larger values of carrier steepness are required to generate instability, up to the limit $p=1$ when the instability domain contracts to the point $\epsilon = \sqrt{2-\sqrt{2}}.$ Conversely, for a given carrier steepness and increasing mode separation the system will eventually exit the instability domain -- this bifurcation is precisely Case (g) in figure \ref{fig:Phase portraits}, where fixed points at $\eta=1$ coalesce and disappear.

\subsection{Nonlinear stability of a bichromatic wave train}
\label{ssec:Nonlinea stability - bichromatic wave train}

Instabilities of bichromatic wave trains, while less well-known than those for uniform wave trains, have also been studied, for example by \cite{Leblanc2009}, \cite{Badulin1995}, or \cite{Ioualalen1994}. While our setting is somewhat different -- in particular, we have only three unidirectionl modes, in contrast to the four modes used by \cite[Sec.\ 4]{Leblanc2009} to discuss class Ib instabilities -- we can nevertheless obtain the nonlinear evolution of such instabilities by our phase-plane analysis, recalling that a bichromatic wave train occupies the line $\eta=0.$

\begin{figure}
    \centering
    \includegraphics[width=0.7\linewidth]{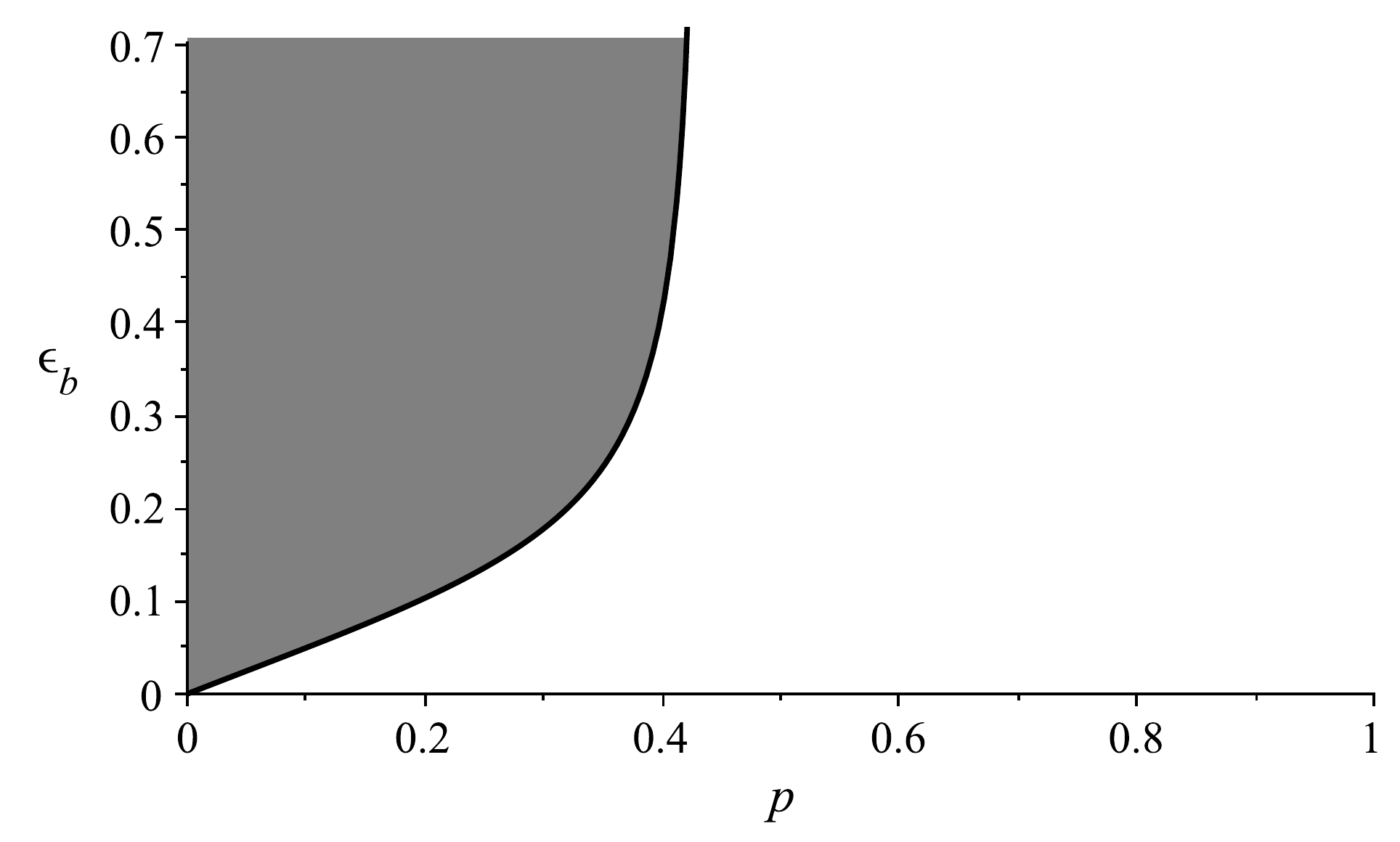}
    \caption{Existence of $\eta=0$ fixed points in $(\epsilon_b,p)$-parameter space, according to \eqref{fixpoint:eta0}. The shaded region denotes the instability domain for a bichromatic wave train with equipartitioned mode amplitudes.}
    \label{fig:eta_0 instability}
\end{figure}

In the present case, where the Fourier amplitudes of the side-bands $k_b$ and $k_c$ are taken to be identical, we naturally have a bichromatic wave train with waves of different steepness. Mode $k_b$ corresponds to a longer wave than mode $k_c,$ and their slopes are related as
\[ \epsilon_b^2 = \left( \frac{k_b}{k_c} \right)^{5/2} \epsilon_c^2, \]
see \cite[Sec.\ 14.6]{Mei2005}. Without loss of generality we depict the instability domain in terms of $\epsilon_b$ in figure \ref{fig:eta_0 instability}.

The existence of fixed points -- and consequent instability -- of the bichromatic wave train is again seen to be generic. For any wave train, there exists a value of mode separation $p$ such that the solution is unstable within the framework of the degenerate quartet. This wave train stabilises for sufficiently large mode separation, as seen in the bifurcation in Case (e) in figure \ref{fig:Phase portraits}. Moreover, we find a critical threshold for mode separation $p\approx 0.42$ beyond which bichromatic wave trains are stable.

\section{Heteroclinic solutions}
\label{sec:Heteroclinic solutions}

The existence of special solutions to the nonlinear Schr\"odinger equation (NLS) has attracted considerable attention in recent years, a discussion of which may be found in \cite{Dysthe1999} or, with an emphasis on hydrodynamics and experimental verification, \cite{Chabchoub2014} and the references therein. We will see that several remarkable solutions -- corresponding to heteroclinic orbits in the phase plane -- exist for the three wave system. These include a primitive breather analogous to that found by \cite{MR915545}, which approaches a plane wave as $t\rightarrow\pm \infty.$ 

The form of the free-surface and its envelope are of particular interest for such solutions; these can be recovered from the solution of the Zakharov equation \eqref{eq: Discrete Zakharov Equation} by defining the following complex amplitude function
\begin{align}
    A(x,t) = \frac{1}{\pi}\left(\sqrt{\frac{\omega_a}{2g}}B_ae^{i(k_ax - \omega_a t)} + \sqrt{\frac{\omega_b}{2g}}B_be^{i(k_bx - \omega_b t)} + \sqrt{\frac{\omega_c}{2g}}B_ce^{i(k_cx - \omega_c t)}\right).
\end{align}
The free surface elevation is then obtained as
\begin{align}
    \zeta(x,t) = \Re\left[A(x,t)\right],
\end{align}
and the envelope is $|A(x,t)|$.

\subsection{Primitive Breather solutions}

The existence of special heteroclinic solutions, with properties similar to the well-known breather solutions of the nonlinear Schr\"odinger equation, clearly depends on the fixed points of our problem.

\begin{figure}
    \centering
    \includegraphics[width=\linewidth]{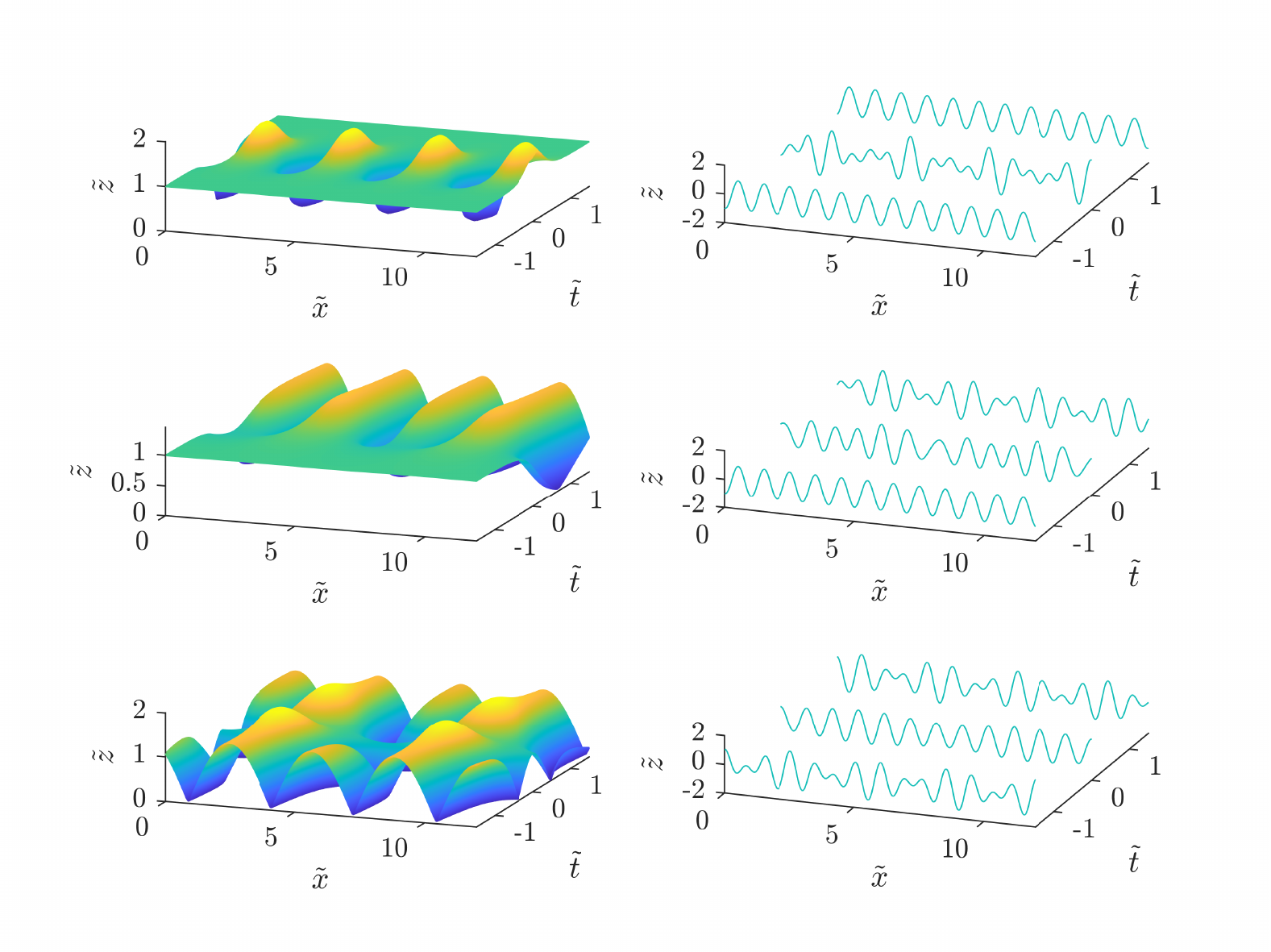}
    \caption{Time evolution of the free surface (right panels) and their envelopes (left panels) along different heteroclinic solutions. In all cases the steepness of the wave $k_a$ is set to $\epsilon_a = 0.2$. Top panels: Primitive Akhmediev breather for $p = 0.3012$. Middle panels: (1-0) breather for the critical $p = 0.1670$. Bottom panels: (0-0) breather for $p = 0.1570$. $\Tilde{x} = x/\lambda_a$, $\Tilde{t} = t\omega_a$ and $\Tilde{z} = z/\epsilon_a.$}
    \label{fig:TimeEvolution}
\end{figure}

\subsubsection{The primitive (1-1) Akhmediev breather}
\label{sssec: (1-1) Breather}
For values of $p$ and $A$ satisfying \eqref{fixpoint:eta1} (see section \ref{ssec:Nonlinear instability - uniform wave train} and figure \ref{fig:eta_1 instability}), the two saddle points  $(\theta,\eta)=(\pm\theta_1,1)$ are connected by a heteroclinic orbit. Along this orbit $\eta<1,$ and a trajectory approaches $(\pm\theta_1,1)$ as $t\to\pm\infty$, respectively. 

In fact, we can compute these solutions explicitly, by considering the following set of equations
\begin{subequations}
\begin{align}
    2\eta\cos(\theta) & = \Delta' + \Omega_0' + \frac{\Omega_1'}{2}\left(1 + \eta\right).\label{Hamiltonian:1-1}\\
    \frac{d\eta}{d\theta} &= \frac{\eta\sin(\theta)}{\cos(\theta) - \frac{\Omega_1}{4}}.\label{detadt:1-1}\\
    \frac{d^2\theta}{dt^2} &= -2\sin(\theta)\frac{d\theta}{dt}.\label{dthetadt:1-1}
\end{align}    
\end{subequations}
Equation \eqref{Hamiltonian:1-1} follows from equating the Hamiltonian \eqref{eq:H:alpha=0} to its value at $\eta = 1$. Equation \eqref{detadt:1-1} is obtained from the implicit function theorem by regarding $\eta$ as a function of $\theta$ instead of $t$. The last equation \eqref{dthetadt:1-1} is obtained by taking the time derivative of equation \eqref{eq:theta:alpha=0}. Equation \eqref{Hamiltonian:1-1} is further used to simplify the form of \eqref{detadt:1-1} and \eqref{dthetadt:1-1}.

Equation \eqref{detadt:1-1} is separable, and can be integrated directly to give
\begin{align} \label{eta:1-1}
    \eta = \frac{4\cos(\theta_1) - \Omega_1}{4\cos(\theta) - \Omega_1},
\end{align}
where the integration constant is chosen so that the limits as $\theta$ tends to $\pm\theta_1$ are 1.

Equation \eqref{dthetadt:1-1} can also be integrated to obtain
\begin{align}
    \frac{d\theta}{dt} = 2\cos(\theta) - 2\cos(\theta_1),
\end{align}
where the integration constant is chosen so that the limits as $\theta$ tends to $\pm\theta_1$ are 0. Integrating again yields an expression for the dynamic phase
\begin{align}\label{theta:1-1}
    \tan(\theta/2) = \tan(\theta_1/2)\tanh(\sin(\theta_1)t),
\end{align}
where the integration constant was chosen so that $\theta(0) = 0.$ Together, \eqref{theta:1-1} and \eqref{eta:1-1} describe the heteroclinic orbits shown in Case (d)--(f) of figure \ref{fig:Phase portraits}. (Note that the heteroclinic orbits around $\theta=\pm \pi$ in Case (b) can be treated similarly.)

The primitive (1-1) Akhmediev breather describes an orbit in phase space which connects one plane-wave solution with another. During the evolution energy is transferred to the side-bands ($\eta$ decreases) while the dynamic phase undergoes little change. Subsequent to this focusing the energy-transfer process reverses, resulting in a phase-shifted plane wave. This situation is depicted in dimensionless coordinates in figure \ref{fig:TimeEvolution} (top panels). The left panel shows the envelope, which is periodic in space and ``breathes'' once at time ${t}=0,$ akin to the Akhmediev breather solution of the NLS. The right panel shows the free surface, which begins as a monochromatic plane wave, grows into a strongly modulated wave-train at  ${t}=0,$ and reverts to a plane wave with an evident phase shift.

\subsubsection{(1-0) breather and (0-1) breathers}
\label{sssec:(1-0) and (0-1) breathers}

A special type of heteroclinic solution appears when the Hamiltonian at $\eta = 1$ vanishes, i.e.\ when $p$ and $A$ are such that $\Delta' + \Omega_0' + \Omega_1'/2 = 0$. There are two such heteroclinic solutions, one linking fixed points $(\theta_0,0)$ and $(\theta_1,1)$ and the other linking $(-\theta_1,1)$ with $(-\theta_0,0)$ (see Case (c) in figure \ref{fig:Phase portraits}).

In this case equation \eqref{eq:eta:alpha=0} simplifies to 
\begin{align}
    \frac{d\eta}{dt} = \pm\eta(1 - \eta)\sqrt{4 - \frac{\Omega_1^2}{4}},
\end{align}
where the sign depends on which fixed point we are considering. Integrating this equation yields 
\begin{align}
    \eta = \frac{e^{\pm\sqrt{4 - \frac{\Omega_1^2}{4}} t + C}}{e^{\pm\sqrt{4 - \frac{\Omega_1^2}{4}} t + C} + 1},
\end{align}
where $C$ is an integration constant depending on the initial conditions. For instance $C = 0$ for the initial condition $\eta(0) = \frac{1}{2}$. 
Assuming that a positive sign is chosen, the solution tends to $1$ as $t\to\infty$ and it tends $0$ as $t\to-\infty$. The limits are reversed if the negative sign is chosen.

The special case described by these orbits corresponds to a complete energy transfer from a uniform wave-train to a bichromatic wave-train (or vice versa). The envelope and free surface are depicted in the middle left and right panels of figure \ref{fig:TimeEvolution}, respectively. These show how spatially-periodic modulation appears ``out of nowhere'' with time ${t},$ or disappears as time is traversed in the negative sense.

\subsubsection{(0-0) breather}
\label{sssec:(0-0)-breather}

For a given (sub-critical) value of mode separation $p,$ and $A$ such that \eqref{fixpoint:eta0} is satisfied (see \ref{ssec:Nonlinea stability - bichromatic wave train} and figure \ref{fig:eta_0 instability}), two distinct fixed points $(\theta,\eta)=(\pm\theta_0,0)$ are linked by a heteroclinic solution that approaches $(\pm\theta_0,0)$ as $t\to\mp\infty,$ and along which $\eta > 0.$

This solution satisfies the following set of equations:
\begin{subequations}
\begin{align}
    2(\eta - 1)\cos(\theta) & = \Delta' + \Omega_0' + \frac{\Omega_1'}{2}\left(1 + \eta\right),\label{Hamiltonian:0-0}\\
    \frac{d\eta}{d\theta} &= \frac{(1 - \eta)\sin(\theta)}{\frac{\Omega_1}{4} - \cos(\theta)},\label{detadt:0-0}\\
    \frac{d^2\theta}{dt^2} &= 2\sin(\theta)\frac{d\theta}{dt}.\label{dthetadt:0-0}
\end{align}    
\end{subequations}
As above, equation \eqref{Hamiltonian:0-0} has been used to simplify equations \eqref{detadt:0-0} and \eqref{dthetadt:0-0}.

Integrating equation \eqref{detadt:0-0} yields
\begin{align}
    \eta = 2\frac{2\cos(\theta) + \Delta + \Omega_0}{4\cos(\theta) - \Omega_1} = \frac{4\cos(\theta) - 4\cos(\theta_0)}{4\cos(\theta) - \Omega_1},
\end{align}
where the integration constant is chosen so that the limits as $\theta$ tends to $\theta_0$ are 0.

Integrating equation \eqref{dthetadt:0-0} yields
\begin{align}
    \frac{d\theta}{dt} = -2\cos(\theta) + 2\cos(\theta_0),
\end{align}
where the integration constant is chosen so the limits as $\theta$ tends to $\pm\theta_0$ are 0. Further integration yields
\begin{align}\label{theta:0-0}
    \tan\left(\theta/2\right) = -\tan(\theta_0/2)\tanh\left(\sin(\theta_0) t\right).
\end{align}

This solution is the natural counterpart to the primitive Akhmediev breather presented in section \ref{sssec: (1-1) Breather} -- however, rather than tending to a uniform wave train with $t\rightarrow\pm \infty$ it tends to a bichromatic wave train, again with an attendant phase shift. The envelope and free-surface for such a case are shown in the bottom left and right panels of figure \ref{fig:TimeEvolution}, respectively. At ${t}=0,$ the modulation is at a minimum, and the free surface elevation is close to a uniform wave-train. 

\subsection{Limiting solutions}

It is clear from equation \eqref{eq:H:alpha=0} that the horizontal lines $\eta = 1$ and $\eta = 0$ are level lines of the Hamiltonian. Hence, any solution of equations \eqref{eq:eta:alpha=0} and \eqref{eq:theta:alpha=0} with suitable initial conditions so that $\eta = 1$ or $\eta = 0$ will remain on these lines.

As with the fixed points themselves, there is no energy exchange among the wave-modes for these solutions. However, they are not steady (or stationary) solutions in the usual sense, because $\theta$ still satisfies the following differential equation:
\begin{align}
    \frac{d\theta}{dt} = 
    \begin{cases}
        \Delta + A\Omega_0 +A\Omega_1 - 2AT_{aabc}\cos(\theta),\quad\text{for $\eta = 1$,}\\
        \Delta + A\Omega_0 + 2AT_{aabc}\cos(\theta),\quad\text{for $\eta = 0$.}
    \end{cases} 
\end{align}
A further differentiation of these equations with respect to $t$ yields equations \eqref{dthetadt:1-1} and \eqref{dthetadt:0-0} respectively, which means that equations \eqref{theta:1-1} and \eqref{theta:0-0} are solutions for each case.
Both limiting configurations of the system exhibit what is sometimes referred to as phase-locking or phase coherence; whereby regardless of the initial phase the dynamic phase converges to the fixed value $-\theta_1$ at $\eta = 1$ and $\theta_0$ at $\eta = 0$. It is expected that if initial conditions are taken close to $\eta = 1$, i.e. if some small amount of energy is initially  put in the side bands, then a similar behaviour will be observed in the evolution of the combined phase, as has been recently reported by \cite{Houtani_2022}.

From the mathematical point of view both limiting configurations $\eta=1$ and $\eta=0$ of the system can be understood as the limit of the generic periodic solutions, as all the energy goes to either the wave $k_a$, or is equipartitioned among the side bands $k_b$ and $k_c$, respectively. From the physical point of view the dynamic phase $\theta$ disappears at both $\eta = 1$ or $\eta = 0$ configurations and one is left with a classical monochromatic Stokes waves or two co-propagating Stokes waves, respectively.

\section{Discussion and conclusions}
\label{sec:Discussion}

The reformulation of the discrete Zakharov equation for three modes as a two-dimensional Hamiltonian dynamical system provides a new and powerful perspective on a classical problem.
In fact, the possibility of reformulating the Zakharov equations for a degenerate quartet in terms of only two auxiliary variables is intimately connected to the existence of an exact solution for this configuration, which can be given in terms of elliptic functions (see \cite{Shemer1985}). With subsequent normalisations, it is possible to effectively describe the entire dynamics in terms of the total wave action $A$ and a single bifurcation parameter $p,$ the mode separation.

The mode separation governs the appearance of fixed points in the phase plane. Saddle points at $\eta=1$ and $\eta=0$ correspond to cases where the entire energy is concentrated in one or two Fourier modes, respectively; the centres correspond to a degenerate quartet of nearly resonant waves which undergoes no time evolution. These fixed points are simply another case of the steady-state waves recently discovered by \cite{Liao2016,Yang_2022}. Our phase-plane analysis provides a simple way to obtain these solutions, and highlights for the first time their critical role in organising the overall dynamics.

The existence of fixed-points on the boundaries $\eta=1$ and $\eta=0$ of our phase-space is intimately connected with the stability of uniform and bichromatic wave trains, respectively. When such fixed points are absent, the trajectories progress along nearly horizontal level lines on the cylindrical phase-plane: the dynamic phase changes, but there is little redistribution of energy. Loosely speaking, we may say that the trajectories do not need to avoid the separatrices connecting the fixed points. The stability results we obtain for uniform wave-trains demonstrate the robustness of the Benjamin-Feir condition, which we recover from the fully nonlinear system without any small-amplitude assumptions. For bichromatic wave trains our setting appears to be different from those previously studied, and no comparable linear results appear to be available.

We also obtain numerous special solutions -- including primitive breathers -- which are identified with heteroclinic orbits in phase space. While the corresponding solutions of the nonlinear Schr\"odinger equation involve a continuum of modes, and therefore exhibit quantitatively different features, the two-dimensional phase space allows for an incisive and transparent analysis. It should be noted that, although breathers have been identified by \cite{Ablowitz1990} with homoclinic orbits, this terminology obscures the phase shift which is apparent in the heteroclinic connections between the saddle points. 

The primitive breather solutions we obtain include an analogue of the Akhmediev breather, which arises from a plane-wave background, as well as limiting cases in which a uniform wave train becomes (as $t\rightarrow \infty$) a bichromatic wave train, and vice versa. While primitive Akhmediev breathers exist in large ranges of parameter space (and thus for many configurations of total energy and mode separation), the limiting (0,1) and (1,0) breathers are unique for given $A$ or $p.$ We also find a seemingly new type of breather-like solution, which arises from, and returns to, a bichromatic background. The maximal energy exchange for this (0,0)-breather is therefore a ``demodulation'', in which the envelope flattens out.

\appendix
\section{Interaction kernels}
\label{app:kernel}
The interaction kernels appearing in the Zakharov equation \eqref{eq:3 Wave-ZE} are lengthy, and expressions for these can be found in \cite{Mei2005} or \cite{Krasitskii1994}. For the case of unidirectional waves in deep water ($k_i>0$ for all $i$) treated in the present study, considerable simplifications to the kernels are possible.
\begin{align}
    T(k,k,k,k)&=\frac{k^3}{4 \pi^2}\\
    T(k_a,k_b,k_a,k_b)&=\frac{1}{4 \pi^2} k_a k_b \min(k_a,k_b)\\ \nonumber
    T(k_a,k_b,k_c,k_d)&=\frac{(k_a k_b k_c k_d)^{1/4}}{32 \pi^2} \left[ \left(k_a k_b\right)^{1/2} + \left( k_c k_d \right)^{1/2} \right]\\
    &\cdot \left( k_a + k_b + k_c + k_d - \left[ |k_a - k_c| + |k_a-k_d| + |k_b-k_c| + |k_b-k_d| \right] \right)
\end{align}
The equation for $T_{abcd}$ is taken from \cite{Kachulin2019} and adjusted by a factor of $2 \pi$. The kernels can be further simplified by making use of the homogeneity property $T(\alpha k_a, \alpha k_b, \alpha k_c, \alpha k_d) = \alpha^{3} T(k_a,k_b,k_c,k_d).$ For a degenerate quartet of waves $2k_a=k_b + k_c$ this enables us to write the kernel
\begin{equation}
    T(k_a,k_a,k_b,k_c)=\frac{k_a^3}{8 \pi^2} \left[ (1-q)(1-q^2)^{1/4}(1+\sqrt{1-q^2})\right],
\end{equation}
where $q=p/|k_a|$ and $p$ is the mode separation distance, such that $k_b = k_a + p, \, k_c = k_a - p.$

\end{document}